\documentclass[prd,showpacs,preprintnumbers,floatfix]{revtex4}
\usepackage{graphicx}
\usepackage{dcolumn}
\usepackage{bm}
\usepackage{amssymb}
\usepackage{epsfig}
\usepackage{color}

\newcommand{\ba}{\begin{eqnarray}}
\newcommand{\ea}{\end{eqnarray}}
\newcommand{\be}{\begin{equation}}
\newcommand{\ee}{\end{equation}}
\newcommand{\bdisplay}{\begin{displaymath}}
\newcommand{\edisplay}{\end{displaymath}}

\input{tcilatex}
\begin{document}

\title{Sensitivity of medium-baseline reactor neutrino mass-hierarchy
experiments to nonstandard interactions}
\author{Amir N. Khan}
\email{amir_nawaz@comsats.edu.pk}
\affiliation{Department of Physics, COMSATS IIT, Park Road, Islamabad, 44000, Pakistan}
\author{Douglas W. McKay}
\email{dmckay@ku.edu}
\affiliation{Department of Physics and Astronomy, University of Kansas, Lawrence, KS 66045}
\author{F. Tahir}
\email{farida_tahir@comsats.edu.pk}
\affiliation{Department of Physics, COMSATS IIT, Park Road, Islamabad 44000, Pakistan}

\begin{abstract}
We explore the impact of nonstandard interactions at source and detector on
the interpretation of reactor electron neutrino disappearance experiments
with short and medium baseline designs. We use the constraints from the
recent results from short baseline experiments and generalize current
estimates of medium baseline event rates to include charged current
interactions at source and detector with standard Lorentz structure but with
nonstandard flavor structure. We find that the average spectrum of observed
events at a baseline of 50 km, in the middle of the currently favored
region, provides a probe of new interactions. We show that an improvement in
sensitivity to nonstandard interactions is possible if combined with improved precision
of input mixing parameters in independent experiments, despite ambiguity
in interpretation of medium baseline data. 
We show that nonstandard
interactions can enhance or suppress the sensitivity of experiments to the
mass hierarchy, depending on the phases of the parameters and the
CP-violating phase in the standard three-neutrino mixing picture.
\end{abstract}

\date{\today}
\maketitle

\section{Introduction}

In the past year the Double Chooz \cite{dchooz}, Daya Bay \cite{db} and RENO 
\cite{reno} electron neutrino disappearance experiments have announced
measurements of the elusive parameter $\sin ^{2}(2\theta _{13})$ in the
vicinity of 0.1. The significance of the measurements of Daya Bay and RENO
is many standard deviations from zero, making their results a milestone in
the understanding of neutrino physics. Daya Bay's result is especially
interesting in that its systematic error is a factor 3 smaller than its
statistical error in the first announced result and a factor 2 smaller in
the second announcement with greatly increased statistics \cite{db2}.  The
total uncertainty is about $12\%$ and the ultimate uncertainty estimate is 1/3 of
that. The plans for substantial increase in precision of these measurement
and ongoing plans for increased precision from accelerator experiments
beyond MINOS \cite{minos} and T2K \cite{T2K} make this a good time to assess
the impact on searches for new physics, such as those surveyed in Refs. \cite%
{klos}, \cite{oz} and \cite{lmrz}. Short and medium baseline reactor
neutrino experiments involve only vacuum oscillation and thus are ideal for
the purpose of revealing effects of flavor changing nonstandard interactions
(NSI) at source and detector \cite{oz}, since flavor conserving NSI and
their contribution to matter effects only play a significant part in
analysis of higher energy, longer baseline experiments.

Though charged lepton decays put stringent bounds on many lepton flavor
violating parameters in the "charged current" modes, the direct evidence
against lepton flavor violation in neutrino experiments, referred to as
"model independent limits", is much weaker \cite{bbf-m}. There is now a
strong effort to use the large value of $\sin ^{2}(2\theta _{13}$) to
determine the mass hierarchy (MH) \cite{pp1,pp2,ldps,mnpf,zwcw,
qetal1,cez1,gp,cez2,qetal2,ghot,cez3} in medium baseline experiments, such
as the JUNO  \cite{DBII, lcwz,yw} and RENO-50 \cite{r50} projects. This goal
was inaccessible in the KamLand experiment \cite{KL}, both because the value
of $\sin ^{2}(2\theta _{13})$ was unknown and the statistics, though
sufficient for measuring $\tan ^{2}\theta _{12}$ and $\Delta m_{21}^{2}$,
were far short of measuring $\sin ^{2}(2\theta _{13})$ and determining the
MH. It is an ideal time to combine the short baseline measurements
of the value of $\sin ^{2}(2\theta _{13})$ and the planned reactor medium baseline,
precision MH measurements to assess the prospects of
seeing new physics. The current status of NSI, including recent work on
reactor neutrinos, is reviewed in \cite{to}.

We implement the short baseline constraint between $\sin ^{2}(2\theta _{13})$
and NSI parameters inherent in "same physics at source and detector", left
handed NSI to determine the conditions required in the medium baseline
experiments to probe deeper into the NSI parameter space to look for new
physics effects. Allowing the input value of $\sin ^{2}(2\theta _{12})$ to
vary, we find a degeneracy between the choice of this mixing angle and the
choice of NSI parameter. Given the current precision of its measurement, we
find this degeneracy limits the reach of the reactor data for probing NSI to
values comparable to those already achieved \cite{bbf-m}.  Higher sensitivity
can be achieved as solar neutrino experiments with different physics in the source, propagation 
and detection chain improve precision in step with that in reactor neutrino
measurements \cite{solaccel}.

Turning to the MH question, we find that the
discrimination between the normal hierarchy (NH) and inverted hierarchy (IH)
of neutrino mass splittings is complicated in the presence of NSI, showing
that the sensitivity to the MH can be significantly affected depending on
the magnitude and sign of an NSI parameter, enhanced if positive and
suppressed if negative.

In the next section we briefly review our formalism and notation, then turn
to a section on short baseline and medium baseline applications and energy
spectra. We follow with a survey of our results on statistical sensitivity
of features of the spectra to NSI and then summarize and conclude.

\section{Formalism of source and detector NSI consequences for neutrino
oscillations}

As pointed out \cite{grossman} and developed \cite{jm1, gb1, jm2, jm3, gb2}
a number of years ago, the presence of lepton flavor violation at the source
and detector of a neutrino beam can skew the
interpretation of neutrino oscillation experiments. For example, the wrong
flavor neutrino provided at the source oscillates and can provide a right
flavor lepton signal at the detector, confusing a wrong signal "appearance"
search, or a wrong signal at the detector, confusing a right
signal``disappearance search". To begin, let us establish our notation by
defining the effective four-fermion, charged current semileptonic Lagrangian
appropriate for the reactor neutrino application. We restrict ourselves to
the case of left-handed neutrino helicity currents and vector and axial
vector quark currents and write \cite{jm2} 
\begin{equation}
\mathcal{L}^s=2\sqrt{2}G_FK_{ij}(\bar{l}_i \gamma_{\lambda}P_LU_{ja}\nu_a)[%
\bar{d}\gamma^{\lambda}( P_L+\beta P_R)u]^{\dagger} +H.c.,  \label{eq:ccLs}
\end{equation}
where indices $i,j,k$ run over flavor basis labels and $a,b,c$ over mass
basis labels, and repeated indices are summed over \cite{fn1}.
The flavor label correspondences are $e$=1, $\mu$=2 and $\tau$=3 and the d and u
spinor fields designate down and up quarks. The coefficients $K_{ij}$
represent the relative coupling strengths for the various lepton flavor
combinations. For reactor disappearance search applications, a nuclear decay
provides the source of neutrinos and the inverse beta decay reaction
provides the electron signal for the detector. From the form of Eq. (\ref%
{eq:ccLs}) it is apparent that the effect of the NSI, represented by the
elements of the dimensionless matrix $K_{ij}$, are captured by the
replacement $U_{ia} \rightarrow K_{il}U_{la}$ in the weak Lagrangian. In the
expression for the oscillation propagation amplitude for antineutrinos of
flavor $i$ at the source to produce leptons of flavor $j$ at the detector,
this amounts to the replacement 
\begin{equation}
\bar{A}_{ij}=U_{ia}e^{-im_a^2\frac{L}{2E}}U^*_{ja}\rightarrow
K_{ik}U_{ka}e^{-im_a^2\frac{L}{2E}}K^*_{jl}U^*_{la}.  \label{eq:nubar}
\end{equation}
Here the neutrino mass eigenvalues are $m_a$, the baseline is $L$, the
propagating neutrino energy is $E$ and, in the present study, we simplify to
the case $\beta$ = 0.  The repeated indices are always taken to be summed. The
corresponding expression for the neutrino beam case reads 
\begin{equation}
A_{ij}=U^*_{ia}e^{-im_a^2\frac{L}{2E}}U_{ja}\rightarrow
K^*_{ik}U^*_{ka}e^{-im_a^2\frac{L}{2E}}K_{jl}U_{la}.  \label{eq:nu}
\end{equation}
In matrix form, the expressions to the right of the arrows in Eq. (\ref%
{eq:nubar}) and Eq. (\ref{eq:nu}) can be written compactly as 
\begin{equation}
\bar{A}=(KU)X(KU)^{\dagger}=K(UXU^{\dagger})K^{\dagger},\; A=\bar{A}^T;
\label{eq:Pmatrix}
\end{equation}
the flavor violating interactions act as a "$K$ transformation" on the
standard oscillation probability. In Eq. \ref{eq:Pmatrix}, the diagonal
matrix $X$ is defined as $X\equiv$ diag$(exp(-2ix_1),\,exp(-2ix_2),%
\,exp(-2ix_3))$, and $2x_a\equiv m_a^2L/2E$. When $X=1$, the unit matrix,
then $\bar{A}=KK^{\dagger} \neq 1$ in general. As we describe later, in the
case of $\bar{\nu_e} \rightarrow \bar{\nu_e}$ in the approximations we adopt
here, $(KK^{\dagger})_{ee}=1$.

\subsection{Effect of source and detector lepton flavor violation on reactor 
$\bar{\protect\nu}_e$ disappearance probability formulas}

To focus on flavor changing NSI, we develop our case with the flavor
changing coefficients $K_{e\mu }$ and $K_{e\tau }$. Because the freedom to
redefine the phases of coefficients is already exhausted by redefining
fermion fields in the standard mixing matrix definition, the elements of the
matrix $K$ are complex in general, and we write $K_{e\mu }=|K_{e\mu
}|exp(i\phi _{e\mu })$ and $K_{e\tau }=|K_{e\tau }|exp(i\phi _{e\tau })$.
Our electron antineutrino propagation amplitude now reads, taking $K_{ee}=1$%
, 
\begin{equation}
\bar{A}_{ee}=(U_{ea}+|K_{e\mu }|e^{i\phi _{e\mu }}U_{\mu a}+|K_{e\tau
}|e^{i\phi _{e\tau }}U_{\tau a})e^{-2ix_{a}}(U_{ea}^{\ast }+|K_{e\mu
}|e^{-i\phi _{e\mu }}U_{\mu a}^{\ast }+|K_{e\tau }|e^{-i\phi _{e\tau
}}U_{\tau a}^{\ast }).  \label{eq:Aee}
\end{equation}%
The oscillation probability factor, $\bar{P}=\bar{A}^{\ast }\bar{A}$ can now
be computed straightforwardly. We will display only the leading order terms
in $K$s in our explicit formulas below, since the flavor violating
coefficients are constrained by experimental searches to be of order 0.05 or
less \cite{bbf-m}. The results we quote are valid to accuracies better than
experimental uncertainties so long as no special choices between the
standard mixing, CP-violating phase $\delta $ and the NSI phases reduce the
linear terms to values much less than the absolute magnitudes of the NSI
parameters. Since only the real part of the parameter $K_{ee}$ contributes
to the disappearance probability, and its value is bounded to be more than
an order of magnitude smaller than the flavor violating parameters $K_{e\mu }
$ and $K_{e\tau }~$\cite{bbf-m}, we do not include it in the first order
formulas. Strictly speaking, the terms in $P_{\bar{\nu}_{e}\rightarrow \bar{%
\nu}_{e}}$ should be normalized, but the normalization affects only higher
order terms in the $K_{ij}$ parameters \cite{fn2}.

Sketching the organization of the $\bar{\nu}_{e}$ disappearance propagation
probability with NSI at source and detector, we write the generic form of
the modulus of the propagation amplitude as 
\begin{equation}
|\bar{A}_{ee}|=|A_{11}+A_{21}e^{-2ix_{21}}+A_{31}e^{-2ix_{31}}|,
\label{eq:genAee}
\end{equation}%
where $x_{ij}=$ $x_{i}-x_{j}=$ $\Delta m_{ij}^{2}L/4E$, with $\Delta
m_{ij}^{2}=m_{i}^{2}-m_{j}^{2}$. In Eq. (\ref{eq:genAee}), the quantities $%
A_{ij}$ are all real. Judiciously using double angle formulas for cosines
and sines and the fact that $A_{11}+A_{21}+A_{31}=1$ in our case, one finds
the expression \cite{fn3}
\begin{equation}
|\bar{A}_{ee}|^{2}=\bar{P}_{ee}=1-(P_{21}\sin ^{2}x_{21}+P_{31}\sin
^{2}x_{31}+P_{32}\sin ^{2}x_{32}).  \label{eq:genAprob}
\end{equation}%
In Eq. (\ref{eq:genAprob}) we define $%
P_{21}=4A_{11}A_{21},P_{31}=4A_{11}A_{31}$ and $P_{32}=4A_{21}A_{31}$. Using 
$\Delta m_{32}^{2}=\Delta m_{31}^{2}-\Delta m_{21}^{2}$, we rewrite Eq. (\ref%
{eq:genAprob}) in a form that is more transparent for discussing the MH
question \cite{ghot}, which reads 
\begin{eqnarray}
\bar{P}_{ee} &=&1-[(P_{21}+\cos (2x_{31})P_{32})\sin
^{2}x_{21}+(P_{31}+P_{32})\sin ^{2}x_{31}-  \nonumber \\
&&\frac{1}{2}P_{32}\sin (2x_{21})\sin (2x_{31})],  \label{eq:genAprob2}
\end{eqnarray}%
where the last term is sensitive to the sign of $x_{31}$ and potentially
provides a handle on the MH \cite{fn4}.
To the approximation we are working in the NSI formalism, Eqs. (\ref{eq:Aee}%
) and (\ref{eq:genAee}) lead to the identifications 
\begin{equation}
A_{11}=c_{13}^{2}c_{12}^{2}-c_{13}\sin (2\theta
_{12})c_{23}K_{-}-c_{12}^{2}\sin (2\theta _{13})c_{23}K_{+},  \label{eq:A0}
\end{equation}%
\begin{equation}
A_{21}=c_{13}^{2}s_{12}^{2}+c_{13}\sin (2\theta
_{12})c_{23}K_{-}-s_{12}^{2}\sin (2\theta _{13})c_{23}K_{+},  \label{eq:A21}
\end{equation}%
\begin{equation}
A_{31}=s_{13}^{2}+\sin (2\theta _{13})c_{23}K_{+},  \label{eq:A31}
\end{equation}%
with the conventions for the standard mixing model (SMM) parameters $%
c_{12}\equiv \cos (\theta _{12})$ etc. as defined in \cite{PDG}. It is
evident from Eqs. (\ref{eq:A0}$-$\ref{eq:A31}) that there are effectively
two NSI parameters in the problem, which we have defined in terms of $%
K_{e\mu }$, $K_{e\tau }$ and mixing parameters and $\delta $, the standard
mixing CP-violating phase, as \cite{fn5}
\begin{eqnarray}
c_{23}K_{+} &\equiv &|K_{e\mu }|\cos (\delta +\phi _{e\mu })s_{23}+|K_{e\tau
}|\cos (\delta +\phi _{e\tau })c_{23}, \\
c_{23}K_{-} &\equiv &|K_{e\mu }|\cos \phi _{e\mu }c_{23}-|K_{e\tau }|\cos
\phi _{e\tau }s_{23}.
\end{eqnarray}%
We will write expressions in terms of these two parameters from now on,
factoring out $c_{23}$ for convenience. This makes explicit
the reduction of the overall strength of the NSI term by a factor $%
c_{23}\simeq s_{23}\simeq 1/\surd {2}$. The coefficients that appear in Eqs. (%
\ref{eq:genAprob}) and (\ref{eq:genAprob2}), namely $P_{21}, P_{31}$ and $P_{32}$, are then given at
first order in the $K$ parameters by the expressions 
\begin{eqnarray}
P_{21}=\sin^2 (2\theta_{12})c^4_{13}+4c_{13}^3\sin(2\theta_{12})\cos(2\theta_{12})c_{23}K_{-}-4c^3_{13}s_{13}\sin^2(2\theta_{12})c_{23}K_{+},
\label{eq:P21}
\end{eqnarray}
\begin{eqnarray}
P_{31}=\sin^2( 2\theta_{13})c^2_{12}-4s^2_{13}c_{13}\sin(2\theta_{12})c_{23}K_{-}+4c^2_{12}\cos(2\theta_{13})\sin(2\theta_{13})c_{23}K_{+},
\label{eq:P31}
\end{eqnarray}
and
\begin{eqnarray}
P_{32}=\sin^2( 2\theta_{13})s^2_{12}+4s^2_{13}c_{13}\sin(2\theta_{12})c_{23}K_{-}+4s^2_{12}\cos(2\theta_{13})\sin(2\theta_{13})c_{23}K_{+}.
\label{eq:P32}
\end{eqnarray}
Taking the parameters $K_{e\mu }$ and $K_{e\tau }$ one at a time, commonly
done in setting NSI bounds, costs little generality. $K_{+}$ and $K_{-}$ are
still independent of each other, because $K_{+}$ depends on the real part of 
$K_{e\mu }$, the imaginary part of $K_{e\mu }$ and the standard mixing phase 
$\delta $, while $K_{-}$ depends only on the real part of $K_{e\mu }$, and
similarly for $K_{e\tau }$. The only loss is the possibility of enhancements
or cancellations that could allow a larger range of possible values than the
quoted one-at-a-time bounds in the literature \cite{bbf-m}.

\section{Combining NSI effects in short and medium baseline experiments}

\subsection{ NSI and the short baseline reactor neutrino determination of $%
\sin^2(2\protect\theta_{13})_{eff}=P_{31}+P_{32}$}

\begin{figure}[tbh]
\begin{center}
\includegraphics[width=4in]{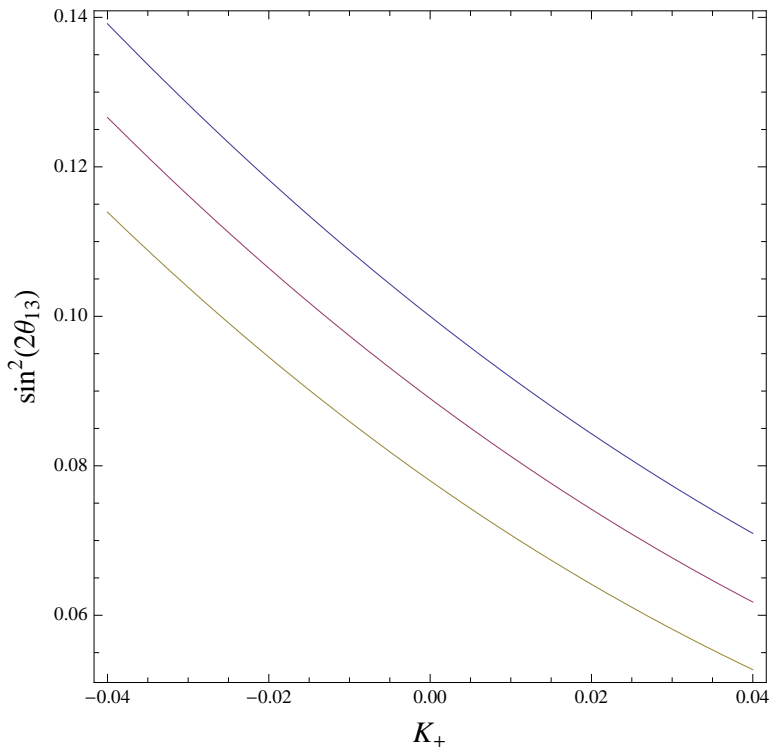}
\end{center}
\caption{The Daya Bay result for $\sin ^{2}(2\protect\theta _{13})_{eff}$
and its one sigma uncertainties \protect\cite{db2} are used to constrain the
ranges of the parameters $\sin ^{2}(2\protect\theta _{13})$ and the NSI
parameter $K_{+}$ in Eq. (\protect\ref{eq:main}).}
\label{fig:constrain1}
\end{figure}

In this case, the term proportional to $\sin ^{2}(x_{31})$ in the final form
of Eq. (\ref{eq:genAprob2}) dominates the contributions to the short
baseline Double Chooz, Daya Bay and RENO experiments. Corrections from $%
x_{21}=\Delta m_{21}^{2}L/4E$ are much smaller than the experimental errors,
so to an excellent approximation the transition probability including the
NSI effects reads 
\begin{eqnarray}
\bar{P}_{ee}=1-\sin^2(\Delta m^2_{31}\frac{L}{4E})(P_{31}+P_{32})=1-\sin^2(\Delta m^2_{31}\frac{L}{4E})[\sin^2(2\theta_{13})+4\cos(2\theta_{13})\sin(2\theta_{13})c_{23}K_+],
\label{eq:main}
\end{eqnarray}
where the angles $\theta _{13}$ and $\theta _{23}$ are the parameterization
angles in the commonly used form of the neutrino mixing matrix \cite{PDG},
and $K_{+}$ is defined in Eq. $(12)$. Only the leading order terms in $K$s
are kept in Eq. (\ref{eq:main}), since the flavor violating coefficients are
constrained by experimental searches to be of order $0.05$ or less \cite%
{bbf-m}. The fit to data by the Double Chooz, Daya Bay and RENO experiments
essentially determine value of the coefficient of $\sin ^{2}(\Delta
m_{31}^{2}\frac{L}{2E})$, which in the standard neutrino mixing picture is
simply $\sin ^{2}(2\theta _{13})$. When the NSI are included, the measured
coefficient of $\sin ^{2}(\Delta m_{31}^{2}L/4E)$ determines the whole
expression in brackets in Eq. \ref{eq:main}. For example, using the Daya Bay
value $\sin ^{2}(2\theta _{13})_{eff}=0.089\pm 0.011$ \cite{db2} we have 
\begin{equation}
\sin ^{2}(2\theta _{13})_{eff}\equiv \sin ^{2}(2\theta _{13})+4\cos (2\theta
_{13})\sin (2\theta _{13})c_{23}K_{+}=0.089\pm 0.011,  \label{eq:constraint}
\end{equation}%
in our linear approximation \cite{fn6}.
We show in Fig. \ref{fig:constrain1} the curve, with corresponding upper and
lower $90\%\ C.L.$ uncertainties, of values in the $(K_{+},\sin ^{2}(2\theta
_{13}))$ plane that satisfy the constraint in the interval $-0.04<K_{+}<0.04$%
. Not all of the parameter ranges quoted in \cite{bbf-m} are
covered in this interval if one allows for maximal constructive coherence
among the parameters, stretching the interval to roughly $-0.1<K_{+}<0.1$,
but it is safely within the region where a linear NSI approximation is
reliable. 

For this application the effective NSI factor is $K_{+}$= $%
|K_{e\mu }|\cos (\phi _{e\mu }+\delta )+|K_{e\tau }|\cos (\phi _{e\tau
}+\delta )$), where we have assumed maximal mixing in the $"23"$ sector, $%
s_{23}=c_{23}=0.717$ \cite{fn7}.  As is known \cite{lmrz}, with the same NSI 
at source and detector, as in our
treatment, the NSI effect an overall shift in the coefficient of the
dominant $\sin ^{2}(x_{31})$ oscillation factor in $\bar{P}_{ee}$. This
establishes a strong correlation between the value of $\sin ^{2}(2\theta
_{13})$ and the value of the NSI parameter $K_{+}$, a point that will be
reiterated in our study of the sensitivity of medium baseline oscillations
to the hierarchy question. To resolve $K_+$ to find evidence for NSI 
requires a combined analysis with an independent,
precision measurement of $\sin^{2}(2\theta _{13})$, involving different 
physics from the physics of reactor $\bar{\nu _{e}}$ disappearance. 
As remarked in Ref. \cite{oz}, this role is played by accelerator 
neutrino appearance experiments, a role which can be filled
 by the ongoing
T2K \cite{T2K} and MINOS \cite{minos} experiments, upcoming experiments 
such as NOvA \cite{nova} and future experiments at a neutrino factory
 \cite{nufac}.

\subsection{Short baseline NSI constraint and medium baseline determination of $P_{21}$ and $P_{32}$}

As the length of the baseline $L$ is increased in the oscillation factor
arguments $x_{21}$ and $x_{31}$, the first and third terms in brackets in
Eq. (\ref{eq:genAprob2}) play the primary role in determining the fraction
of $\bar{\nu}_{e}s$ surviving at distance $L$ from the source. The first
term's $P_{21}$ coefficient controls the "slow" oscillations with energy at fixed baseline, while
the third term's coefficient, $P_{32}$, controls the "fast" oscillations. Solving Eq. (\ref{eq:constraint}) 
for $\sin ^{2}(2\theta _{13})$ in terms of $K_{+}$,  as illustrated in Fig. \ref{fig:constrain1}, 
we can write the $x_{ij}$%
-independent factors in the probability for $\bar{\nu}_{e}$ disappearance,
Eq. (\ref{eq:genAprob2}), completely in terms of the NSI parameters $K_{+}$
and $K_{-}$ and the standard mixing angles $\theta _{12}$ and $\theta _{23}$. 
Alternatively, we can eliminate $K_+$ in terms of $\theta_{13}$ and
$\theta_{23}$, as we do in Sec. III.B.2 below. 
First we take a look at the overall spectral behavior and then turn to a
quick survey of the rapid oscillations that contain the information about
the MH.

\subsubsection{Study of a modeled medium baseline reactor neutrino
experiment and the influence of NSI effects}

Here we follow several recent explorations of the prospects for MH
determination within standard model extended by mixed, massive neutrinos.
The favored baselines are in the neighborhood of (30-60) km \cite{qetal1}, \cite{qetal2}
and \cite{ghot}. As in \cite{ghot} we adopt the neutrino
flux model of \cite{ve}, the gaussian energy resolution smearing model of 
\cite{ghot}, the cross section of \cite{vb} and, the general approach of 
\cite{ghot}. The observed neutrino energy distribution for an experiment
with a $20$ giga watt (GW) thermal power, a detector of 5 kilotons fiducial
volume and $12\%$ weight fraction of free protons with total number of free
protons $N_{p}$ and $5$ years of exposure time $T$, can be written

\begin{equation}
\frac{dN}{dE_{\nu }}(E_{\nu })=\frac{N_{p}T}{4\pi L^{2}}%
\int_{E_{th}}^{E_{max}}dE\frac{dN}{dE}\bar{P}_{ee}(L,E,K_{+},K_{-})\sigma
_{IBD}(E)G(E-E_{\nu }).  \label{eq:spect}
\end{equation}%
Here $E_{th}=m_{n}-m_{p}+m_{e}$, the threshold energy, $\frac{dN}{dE}$ is
the rate of neutrino emission from the reactor per $MeV$, $\bar{P}_{ee}$ is
defined in Eqs. (\ref{eq:genAprob2}), (\ref{eq:P21}), (\ref{eq:P31}) and (%
\ref{eq:P32}), $\sigma _{IBD}$ is the total cross section for inverse beta
decay, and $G(E-E_{\nu }$) is the gaussian smearing function that takes into
account the response of the detector to the deposited energy.

\begin{equation}
G(E-E^{\prime },\sigma _{E}(a,b))=\frac{1}{\sqrt{2\pi }\sigma _{E}(a,b)}exp%
\left[ -\frac{(E-E^{\prime})^ 2}{2(\sigma _{E}(a,b))^{2}}\right] .
\label{eq:G}
\end{equation}%
The uncertainty $\sigma _{E}(a,b)$ in the energy $E$ is parameterized in
terms of a statistical parameter $a$ and a systematic parameter $b$ as

\begin{equation}
\sigma _{E}(a,b)=E\left[ \frac{a^{2}}{E}+b^{2}\right] ^{1/2},
\label{eq:sigE}
\end{equation}%
where energy is in $MeV$ and the first term under the square root represents
the statistical uncertainty in the energy deposited in the detector and the
second represents the systematic energy scale uncertainty.

\begin{figure}[tbph]
\begin{center}
\includegraphics[width=6in]{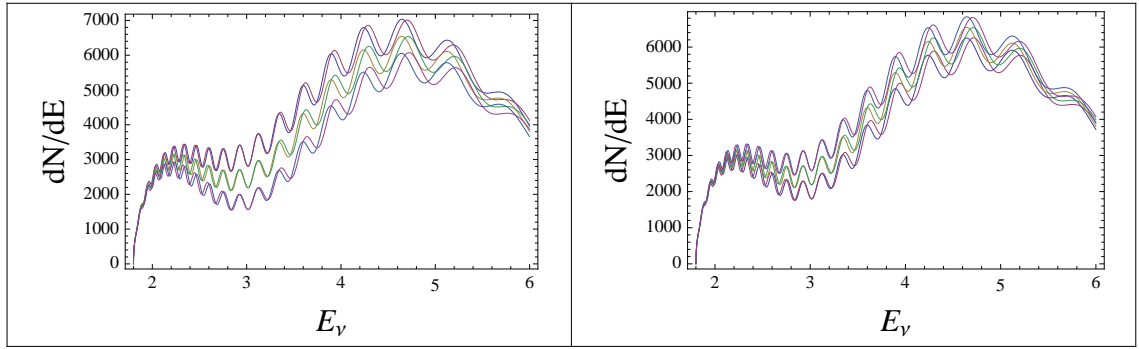}
\end{center}
\caption{The predicted rates of events with a 50 km baseline for the medium
baseline set up described in the text with perfect detector energy response,
a=b=0. $E_{\protect\nu }$ is in MeV and dN/d$E_{\protect\nu }$ is in inverse
MeV. The left plot shows variation of the spectrum with central values of
all parameters and different NSI choices. From top to bottom: $%
K_{+}=-0.04=K_{-}$, where $P_{32}$ is minimal, $K_{+}=0=K_{-}$, 
no NSI, and $K_{+}=-\ 0.04$ and $K_{-}=+\ 0.04$, where $P_{32}$ 
is maximal. NH curves are shown in blue, IH in red. 
The right plot shows the variation of the spectrum with no NSI and 
$\sin^{2}(2\protect\theta _{12})$ ranging within its 1$\protect\sigma $
uncertainty. The middle curves on the left and right are the same.}
\label{fig:rates0}
\end{figure}

Computing the energy spectrum expected in the model described, we show the
results at a baseline of 50 km with perfect detector response (no smearing)
in Fig. \ref{fig:rates0}. Table \ref{tab:inputs} displays values and their
uncertainties for the relevant input parameters.

\begin{table}[htbp]
\begin{center}
\renewcommand{\arraystretch}{1.5} 
\begin{tabular}{|l|l|l|l|r|}
\hline
$\sin^2(2\theta_{12})$ & $\sin^2(2\theta_{13})$ & $\Delta m^2_{21}$ & $%
\Delta m^2_{31}$ &  \\ \hline\hline
$0.857 \pm 0.024^*$ \cite{KL,PDG} & $0.089 \pm 0.011^*$ \cite{db} & $(7. 50
\pm 0.20)10^{-5}$$^*$ \cite{PDG} & $(2.32 \pm 0.10) 10^{-3}$$^*$ \cite{PDG}
&  \\ \hline
$0.869^{+0.031}_{-0.037}$ \cite{SNO} &  &  &  &  \\ \hline
$0.845^{+0.066}_{-0.071}$ \cite{KL} &  &  &  &  \\ \hline
\end{tabular}
\renewcommand{\arraystretch}{1}
\end{center}
\caption{Values of relevant input parameters. The values marked with
asterisks are the ones used in the numerical work to illustrate the ideas
and make estimates. }
\label{tab:inputs}
\end{table}
The left plot shows variation of the spectrum with central values of all
input parameters and different NSI parameter choices. NH curves are
shown in blue, IH in red. When $x_{21}$ = $\pi $ $/2$ at $E_{\nu }\sim $ $3\
GeV$, the maximum and zero in the factors $\sin ^{2}(x_{21})$ and $\sin
(2x_{21})$, respectively, is evident in the figure. The right plot shows the
variation of the spectrum with no NSI and, from top to bottom, $\sin
^{2}(2\theta _{12})$ at its central value minus $1\sigma $, its central
value and its central value plus $1\sigma $. The middle curves on the left
and right are the same, of course, and are included for reference \cite{fn8}. 
As we discuss in the following
subsection, the ordering is expected, since the survival probability is
smallest and the rates of detection are smallest when the "oscillation
term", which represents transitions into other flavors, is largest within a
given range of parameter space. The parameter choice that makes the
MH-sensitive term, $P_{32}$, largest also makes the overall oscillation term
in Eq. (\ref{eq:main}) take its largest value and makes the survival rate
the smallest. For illustration we have chosen NSI magnitudes that are about
as large as allowed by the current experimental constraints on NSI
parameters when taken one at a time. 

\begin{figure}[htbp]
\begin{center}
\includegraphics[width=6in]{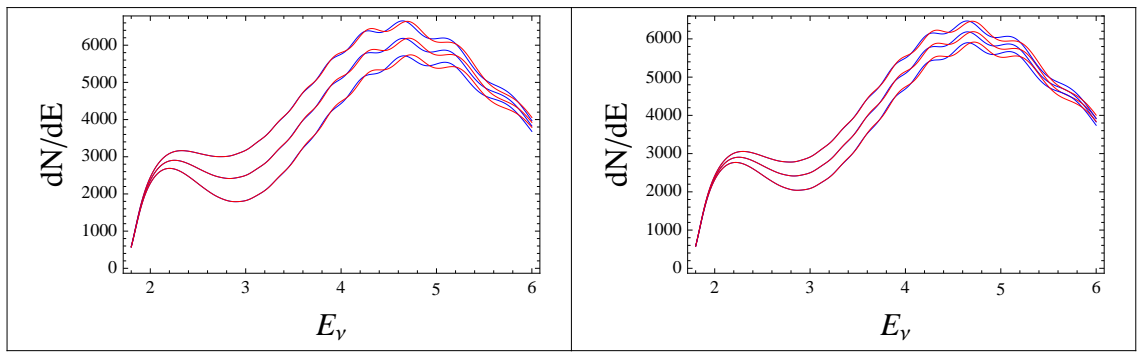}
\end{center}
\caption{The predicted rates of events with a $50~km$ baseline for the
medium baseline set up described in the text with $a=6\%$ and $b=0$
uncertainties in the detector energy response. $E_{\protect\nu }$ is in MeV
and $dN/dE_{\protect\nu }$ is in inverse MeV. The left plot shows variation
of the spectrum with central values of all input parameters and different NSI
choices : top to bottom are $K_+ = K_-$ = -0.04, $K_+=K_-=0$,  
$K_+=-0.04$ and  $K_-=+0.04$. NH curves are shown in blue, IH in red.
The right plot shows the variation of the spectrum with no NSI and $\sin ^{2}(2\protect%
\theta _{12})$ ranging within its $1\protect\sigma$ uncertainty. The middle
curves on the left and right are the same.}
\label{fig:rates6}
\end{figure}

In Fig. \ref{fig:rates6} we show the same cases but with energy uncertainty
fractions $a=6\%$ and$~b=0$. Within this range of NSI possibilities, it is
clear that the effects on the spectrum can be large and mimic those of the
uncertainty in the value of $\sin ^{2}(2\theta _{12})$. If the value of $%
\sin ^{2}(2\theta _{12})$ could be determined with high precision from solar
neutrino data, for example, then a combined analysis with
the medium baseline spectrum could disentangle the input uncertainties and the NSI andprovide
a sensitive probe of the NSI parameters $K_{-}$. In Sec. IV.  we show 
in another way that within the current
uncertainties in $\sin ^{2}(2\theta _{12})$, there is a degeneracy
between $K_{-}$ and $\sin ^{2}(2\theta _{12})$ in interpreting medium
baseline data, as one can see in the expression for $P_{21}$ in Eq. \ref%
{eq:P21}, the dominant coefficient of $\sin ^{2}x_{21}$. Changes in the
value of $\sin ^{2}(2\theta _{12})$ can be compensated by changes in $K_{-}$, %
for example. In terms of Figs. \ref{fig:rates0} and \ref{fig:rates6}, one
finds that the left and right plots are essentially identical if one choses $%
|K_{+}|~=|K_{-}|~=0.024$ and the central value of $\sin ^{2}(2\theta _{12})$
for the left plot. We will return to this issue in Sec. IV. 

\begin{figure}[htbp]
\begin{center}
\includegraphics[width=4in]{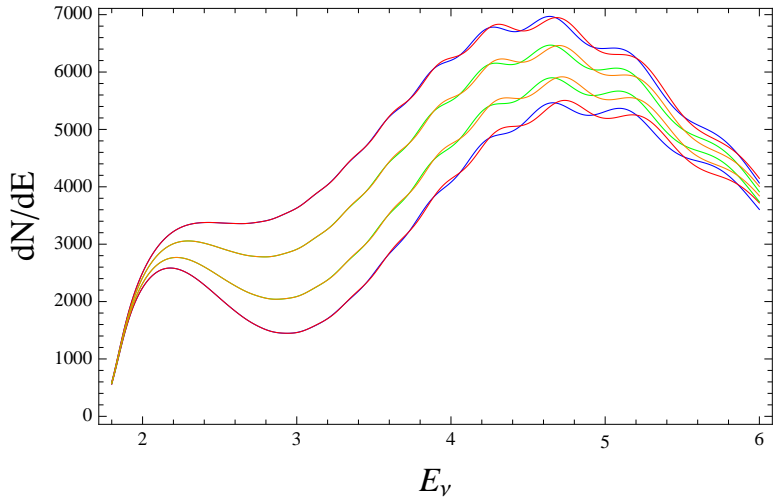}
\end{center}
\caption{Allowing our full range values for the NSI parameters $|K+|$ and $%
|K_-|$ and the mixing angle $\protect\theta_{12}$, we find the outer
envelope of spectral curves indicated in the figure. The envelope of curves
expected just from the 1 $\protect\sigma$ variation of $\protect\theta_{12}$
is shown inside the full envelope. $E_{\protect\nu}$ is in MeV and dN/d$E_{%
\protect\nu}$ is in inverse MeV.}
\label{fig:envelopes}
\end{figure}

In Fig. \ref{fig:envelopes}, we show the envelopes of all the spectra
possible within the parameter ranges for $|K_{+}|$, $|K_{-}|$ and $\sin
^{2}(2\theta _{12})$. It is clear that there is a wide range of possible
spectra beyond that expected from the uncertainty in the input values of the
mixing angles when the NSI effects are included.

The differences in the overall rates are much larger in Figs. \ref%
{fig:rates0} and \ref{fig:rates6} than the purely statistical fluctuations
in counting estimated by the square root of the event numbers, even with the
energy smearing. This fact is behind the expectation of highly improved
precision in determination of the coefficient of $\sin ^{2}x_{21}$ \cite%
{ghot}, $P_{21}$, $\sin ^{2}(2\theta _{12})$ in the SMM. The primary effect
of the smearing is to reduce the distinction between NH and IH expected
spectra, as studied in \cite{ghot}. We show this in a different, more direct
way in Fig. $5.$

\begin{figure}[tbph]
\par
\begin{center}
\includegraphics[width=6in]{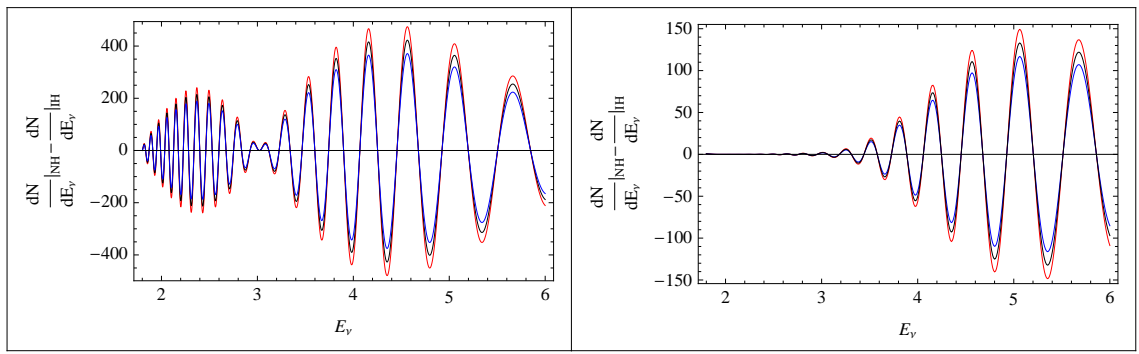}
\end{center}
\caption{The predicted difference between rates of NH and IH events as a
function of energy with a 50 km baseline for the medium baseline set up
described in the text. $E_{\protect\nu }$ is in MeV and $dN/dE_{\protect\nu }
$ is in inverse MeV. The left panel shows the result with no uncertainty in
the detector energy resolution and energy scale and the right shows the
result with $6\%$ uncertainty in the detector energy resolution. From top to
bottom: $K_{+}=-0.04$ and $K_{-}=+0.04$, where $P_{32}$ is maximal, $%
K_{+}=0=K_{-}$, no NSI, and $K_{+}=-0.04=K_{-}$, where $P_{32}$ is minimal.
Input parameters are set at their central values. Note the difference in scale between
the two plots.}
\label{fig:difNHIH}
\end{figure}
Taking the \emph{difference} between the NH rate prediction and the IH
prediction gives a clearer picture of the distinction between the two
possibilities. In Fig. \ref{fig:difNHIH} the left panel shows the difference
between the NH and IH predictions as a function of energy with a baseline of
50 km when the resolution is perfect. For this ideal situation, the
distinction between the NH and IH predictions is clear in the maximal and
no NSI cases but marginal in the minimal case. When the uncertainty in the
energy resolution is $6\%$ and the energy scale error is $0$, the
distinctions are marginal at best in the maximal case and insufficient in
the other cases. In \cite{ghot} this case was made in detail for the SMM,
where the authors concluded that a resolution of better than $3\%$ would be
necessary at the optimal baseline to resolve the hierarchy in a reasonable
time. In the right panel in Fig. \ref{fig:difNHIH}, we show the plot of the
difference between NH and IH rates as a function of energy at 50 km
baseline with a $6\%$ statistical energy resolution uncertainty and no
statistical uncertainty as modeled by the gaussian smearing function. As
shown in the figure, the NSI can cause a shift of up to $12\%$ in the
sensitivity of the spectrum to the MH. With no NSI but allowing for a $%
1\sigma $ variation in the value of $\sin ^{2}(2\theta _{13})$, the results
look nearly the same as shown in Fig. \ref{fig:difNHIH}, reflecting the $12\%
$ uncertainty in $\sin ^{2}(2\theta _{13})_{eff}$. When the anticipated $4\%$
goal for the uncertainty in $\sin ^{2}(2\theta _{13})_{eff}$ is reached, the
sensitivity to NSI effects will be correspondingly enhanced by a factor$~3$.
The uncertainty in $\sin ^{2}(2\theta _{12})$ allows a shift of about $3\%$. 

\subsubsection{The influence of the NSI parameters $K_{+}$ and $K_{-}$ on
the factors $P_{21}$ and $P_{32}$ in the disappearance probability}

The coefficient that controls the scale of long oscillation length behavior
of the disappearance probability is $P_{21}$, which has the signature
feature that it is independent of the NSI parameter $K_{+}$ to a high degree
of accuracy. Over the range $-0.04\leq K_{+}\leq 0.04$, the change in $P_{21}
$ for any choice of $K_{-}$ in the same range is at most $0.1\%$, typically $%
100$ times smaller than the change due to $K_{-}$ over this range. To 
see how this happens, it is helpful to use Eq.(\ref{eq:constraint}) to eliminate $K_+$ 
and rewrite Eq. (\ref{eq:P21}) in the form
\begin{equation}
P_{21}=\sin^{2}(2\theta _{12})c_{13}^{4}\left(1- 4\tan\theta_{13}\frac{(\sin^2(2\theta_{13})_{eff}-\sin^2(2\theta_{13}))}{\sin(2\theta_{13})\cos(2\theta_{13})}\right)+4c_{13}^{3}\sin (2\theta _{12}) \cos(2\theta _{12})c_{23}K_{-}.  \label{eq:P21new}
\end{equation}%
The factor $c_{13}^4$ multiplying $\sin^2(2\theta_{12})$ in the first term increases monotonically as 
$\sin^2(2\theta_{13})$ decreases from its maximum value to its minimum value, while the
factor in the large parentheses decreases monotonically.  The two effects compensate each other to
high accuracy. The whole first term is of order 1.  The coefficient of $K_-$ in the second term increases monotonically by 2\% over the same range.  The magnitude of the coefficient is close to 1, so the size of the second term relative to the first is determined by $K_-$, which ranges between -0.04 and +0.04.  The net effect is that $P_{21}$
varies by at most 0.1\% with $-0.04< K_+ <+0.04$ for any chosen values of $K_-$ and the
input values of mixing angles within their uncertainties.

The dependence of the MH-sensitive coefficient $P_{32}$ on the NSI
parameters can be substantial \cite{fn9}. The value of $P_{32}$ largely
controls the possibility of determining the sign of the last term in Eq. (%
\ref{eq:genAprob2}) and, therefore, whether $m_{3}>m_{1}$ (NH) or $%
m_{1}>m_{3}$ (IH). The effects of new flavor violating interactions on $%
P_{32}$ increase as $K_{+}$ decreases, which corresponds to increasing
values of $\sin ^{2}(2\theta _{13})$, shown in Fig. \ref{fig:constrain1}.
The smallest values of $P_{32}$ occur at the smallest values of $K_{+}$ and $%
K_{-}$. For example, at $(-0.05,-0.05),$ $\frac{1}{2}P_{32}=0.0115$, while
at $(-0.05,+0.05),$ $\frac{1}{2}P_{32}=0.0165$. 
To understand the dependence of $P_{32}$ on the $(K_{+},K_{-})$ parameters, again
it is helpful to use the definition of $\sin ^{2}(2\theta _{13})_{eff}$ in
Eq. (\ref{eq:constraint}) to rewrite $P_{32}$ in an equivalent form as 
\begin{equation}
P_{32}=s_{12}^{2}\sin ^{2}(2\theta _{13})_{eff}+4s_{13}^{2}c_{13}\sin
(2\theta _{12})c_{23}K_{-},  \label{eq:P32new}
\end{equation}%
in which  $P_{32}$ has no \emph{explicit} dependence on $%
K_{+}$. Through the constraint Eq. (\ref{eq:constraint}), $P_{32}$ has \emph{%
implicit} dependence on $K_{+}$ via the factor $s_{13}^{2}c_{13}$ in the
coefficient of $K_{-}$. For fixed $K_{+}$, the value of $P_{32}$ is linearly
dependent on $K_{-}$ with a positive coefficient. Because of the constraint
imposed by Eq. (\ref{eq:constraint}), $\sin (\theta _{13})$ decreases as $%
K_{+}$ increases, which decreases the coefficient of $K_{-}$, and conversely
when $K_{+}$ decreases. In short, $P_{32}$ increases as $K_{+}$ grows when $%
K_{-}<0$ and decreases as $K_{+}$ grows when $K_{-}>0$, though the
dependence is weak compared to the dependence on $K_{-}$.

The upshot is that our parameterization in terms of $K_{-}$ and $K_{+}$ has the good feature
that we can choose to write the coefficients $P_{21}$ and $P_{32}$ with only
 $K_{-}$ involved explicitly in the NSI effects on
overall spectrum shape and, as we argue below, in the sensitivity, or not,
of the MH to NSI. The parameter $K_{+}$ plays an indirect role through its
relationship with the measured $\sin ^{2}(2\theta _{13})_{eff}$ and the
mixing angle $\theta _{13}$, Eq. (\ref{eq:constraint}).  The weak dependence on 
$\sin^2(2\theta_{13})$ is transparent and makes the minor role of $K_+$ easier to 
understand.

A general feature of $P_{32}$, the "contrast coefficient" between the NH and
IH rate values, is that for all \emph{positive} values of $K_{-}$ the
differences between NH and IH rates are larger than those for standard
mixing and the overall rates are smaller, while the opposite is true for all 
\emph{negative} values of $K_{-}$. When $K_{-}=0$, there is a degeneracy
that is almost complete; given an input value of $\sin ^{2}(2\theta
_{23})_{eff}$, Eq. (\ref{eq:constraint}), $P_{32}^{NSI}=P_{32}^{SMM}$ and $%
P_{31}^{NSI}=P_{31}^{SMM}$, and $P_{21}^{NSI}-P_{21}^{SMM}\ll 1$ for all
values of $K_{+}$ within current experimental bounds. Only for values of $%
K_{-}\neq 0$ can the presence of NSI be detected with any confidence. How
large must $K_{-}$ be for its detection to be possible? We address this
question in the following section.

\section{Statistical sensitivity of the spectrum to the NSI parameters}

The discussion in Sec. III.B.2 laid out the main features of
the NSI parameters' influence on $P_{21}$ and $P_{32}$, where the first
controls the long oscillation length behavior of the spectrum and the second
controls the contrast between the NH and IH short oscillation length
behaviors of the spectrum. The examples shown in Figs. \ref{fig:rates0} 
through \ref{fig:difNHIH} are chosen with
values of the NSI parameters suggested by the individual parameter bounds 
\cite{bbf-m} in order to make the effects visually clear. In this section we
make the pictures more quantitative by using simple $\chi ^{2}$ estimates of
the deviation from the SMM when NSI effects are included, while illustrating
the interplay with the input parameter uncertainties by letting key inputs
range over their one standard deviation values.

\subsection{ Fitting SMM "data" with NSI fit function and $\sin^2(2\protect%
\theta_{12})$ "pull"}

As an estimator, we adopt the "no NSI", or SMM, values for the spectrum as
data and the square root of the number of predicted events as the
statistical uncertainty. Using the same points spaced at intervals of $0.01$
MeV from which Figs. \ref{fig:rates0} through \ref{fig:difNHIH} were made,
but with the smearing parameters chosen as $a=0.03\%$ and $b=0$, we compute 
\begin{equation}
\chi ^{2}=\Sigma _{i}\left( \frac{\frac{dN}{dE_{\nu }}^{NSI}-\frac{dN}{%
dE_{\nu }}^{SMM}}{\surd \frac{dN}{dE_{\nu }}^{SMM}}\right) _{i}^{2}(\Delta
E_{\nu })_{i}.  \label{eq:chisq}
\end{equation}%
Input parameters are fixed at central values for $\frac{dN^{SMM}}{dE_{\nu }}$%
, while one or more are allowed to vary over their $1\sigma $ ranges for $%
\frac{dN^{NSI}}{dE_{\nu }}$. Because $\chi ^{2}=0$, its minimum, when NSI
parameters are zero and the input parameters are at their central value, the
value of $\Delta \chi ^{2}\equiv \chi ^{2}-\chi _{min}^{2}$ is the same as $%
\chi ^{2}$, up to a pull "penalty" term that contributes to the $\chi ^{2}$
value for each input parameter that is varied away from its central value. 

\begin{figure}[tbph]
\par
\begin{center}
\includegraphics[width=4in]{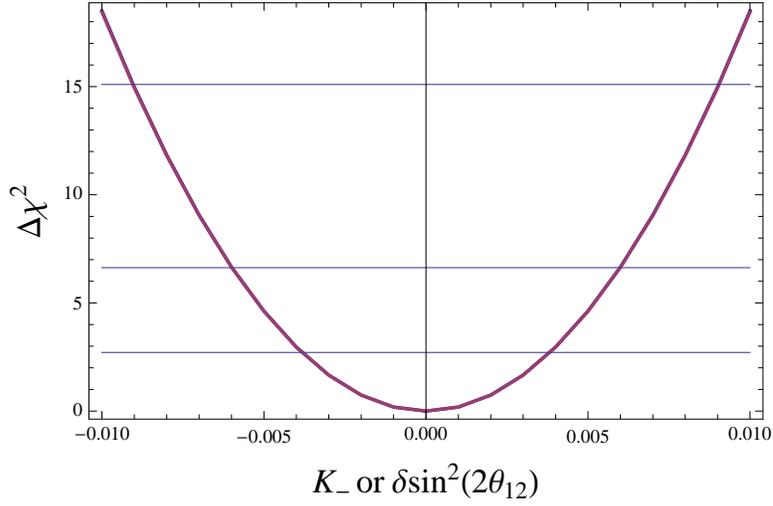} 
\end{center}
\caption{For $K_{+}$ and $K_{-}$ ranging between -0.01 and +0.01, we show
the boundaries of the 90\% C.L., the 99\%C.L. and the 99.99\%C.L.
projected on a constant $K_{+}$ plane. The high degree of degeneracy along
the lines of fixed $K_{-}$, see Fig. \protect\ref{fig:chisqvks}, shows that $%
\Delta \protect\chi ^{2}$ is independent of $K_{+}$ to high accuracy, and
the problem reduces to $1$ parameter. The energy smearing and systematic
parameters chosen for the plots are $a=0.03~$and $b=0,$ respectively, but
the changes in the plot are negligible when any values $a\leq 0.06$ and $%
b\leq 0.01$ are used.}
\label{fig:90etcCL}
\end{figure}

\begin{figure}[tbph]
\begin{center}
\includegraphics[width=4in]{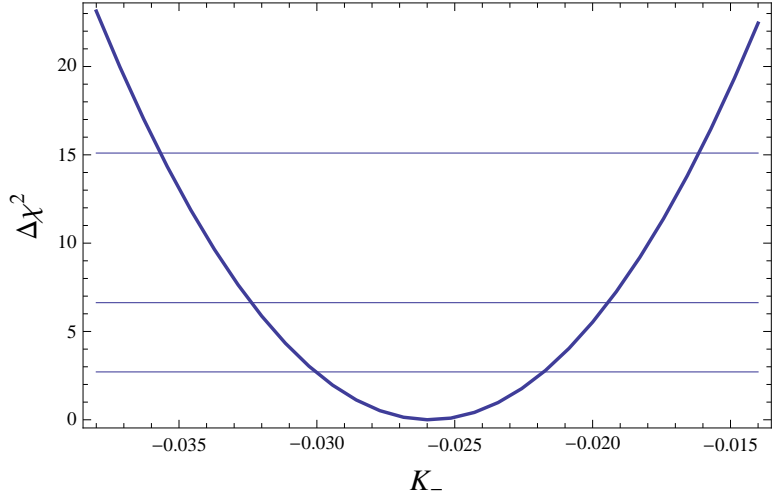}
\end{center}
\caption{The various $C.L.$ limits for a fit to SMM with NSI and $\sin ^{2}(2%
\protect\theta _{12})=0.881$. The other input parameters are set at their
current central values and the energy uncertainty parameters are $a=0.03$
and $b=0,$ while$\ K_{+}=0$ for the plot. The results are not sensitive to
these choices. For clarity of display, the additive $\protect\chi ^{2}=1$
penalty is not included in the plot.}
\label{fig:1dchisqx12}
\end{figure}

\begin{figure}[tbph]
\par
\begin{center}
\includegraphics[width=4in]{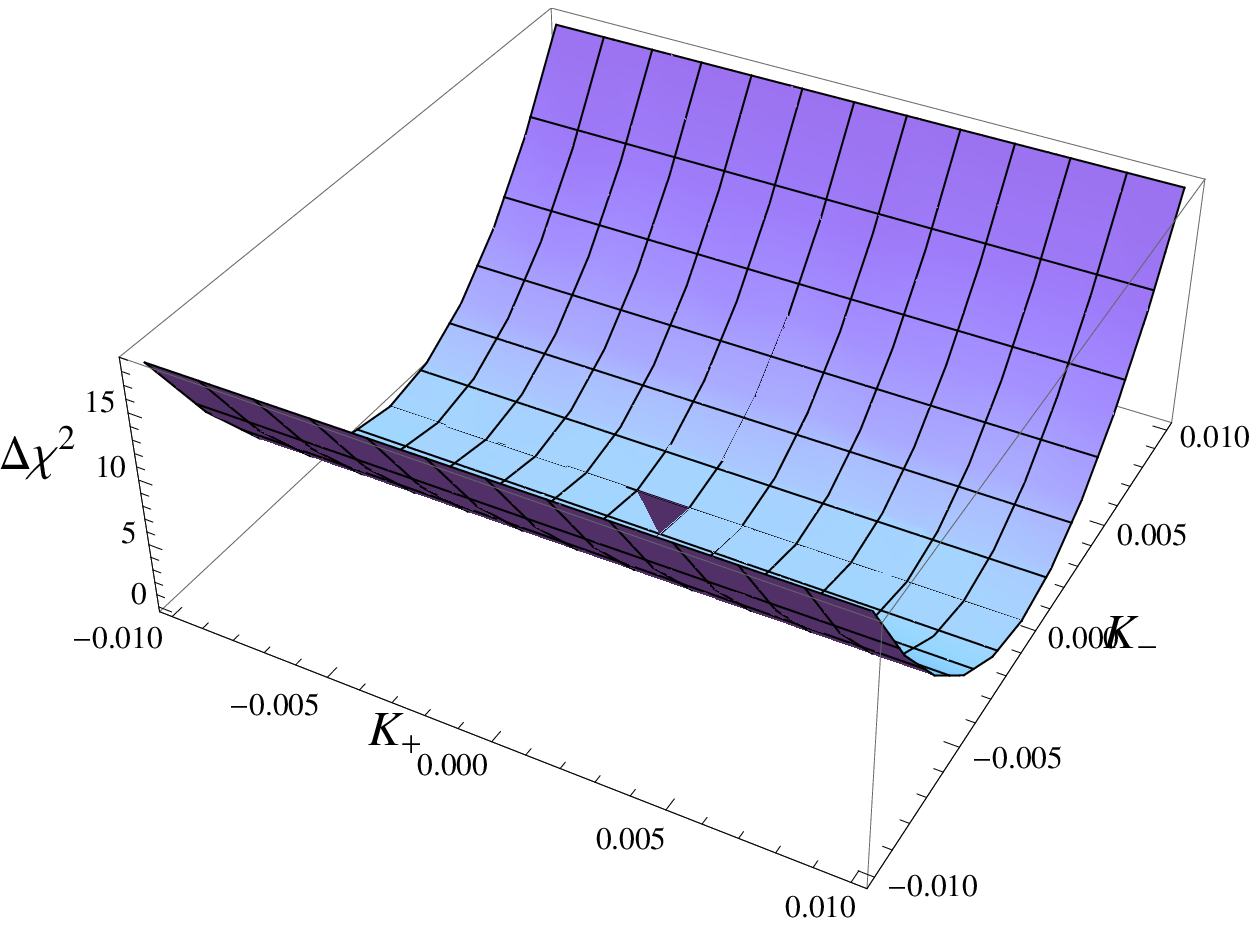} 
\end{center}
\caption{For $K_{+}$ and $K_{-}$ ranging between $-0.01$ and $+0.01$ and the
special case where \emph{all} input parameters are set to their central
values for $\frac{dN^{NSI}}{dE_{\protect\nu }}$, we show the values of $%
\protect\chi ^{2}$ as computed in Eq. (\protect\ref{eq:chisq}). The
degeneracy along the line $K_{-}=0$ is obvious. }
\label{fig:chisqvks}
\end{figure}

Since the spectrum shape at a given medium baseline value is controlled by $%
\Delta m_{21}^{2}$, and $\sin ^{2}(2\theta _{12})$ determines the scale of
the oscillations, we do a statistical analysis allowing the latter to vary
 away from its central value of $\left\langle \sin^{2}(2\theta _{12})\right\rangle =0.857$ 
in the NSI fit function, while the SMM "data" has its input parameters fixed 
at their central values. There is an implicit, additive "pull" term 
$(($ $\sin^{2}(2\theta _{12})$- $\left\langle \sin ^{2}(2\theta _{12})\right\rangle
)/\sigma _{12})^{2}$ in Eq. (\ref{eq:chisq}). In practice we search
for a minimum in $\chi ^{2}$ with $K_{-}$ and $K_{+}$ for a range of values
of $\sin ^{2}(2\theta _{12})$, chosen simply for illustration to be within
its quoted $1\sigma $ uncertainty range, $\sigma _{12}=0.024$. Graphically,
we show the example of the results when $K_{+}=0$ and $\sin ^{2}(2\theta
_{12})=0.857$ in Fig. \ref{fig:90etcCL} and the example when $\sin
^{2}(2\theta _{12})=\left\langle \sin ^{2}(2\theta _{12})\right\rangle
+\sigma _{12}=0.881$ in Fig. \ref{fig:1dchisqx12}. The landscapes of $\chi
^{2}$ values in the $(K_{+},K_{-})~$plane are all similar to that shown in
Fig. \ref{fig:chisqvks} for the case where $\sin ^{2}(2\theta _{12})=0.857$,
with a near degeneracy in $K_{+}$ along fixed $K_{-}$ values. Again, since $%
P_{21}$ is independent of $K_{+}$ to an excellent approximation, the
dependence of $\chi ^{2}$, Eq. (\ref{eq:chisq}) on the value of $K_{+}$ is
negligible and the results are insensitive to the choice of energy
uncertainty parameters. In Fig. \ref{fig:90etcCL}, overlaid is the plot of $\chi^2$
vs. $\delta \sin^2(2\theta_{12})$, the deviation from the central value, with $K_+=K_-=0$.
The two plots are degenerate to a fraction of a percent!

Modulo the $\chi ^{2}=1$ penalty, Fig. \ref{fig:1dchisqx12} shows that the
fit is essentially degenerate with the fit with $\chi ^{2}=0$ at $\sin
^{2}(2\theta _{12})=0.857$ and with values of $K_{-}$ consistent with
current bounds on NSI and $\sin ^{2}(2\theta _{12})$ at the upper end of its
current $1\sigma $ range. As one explores other values of $\sin ^{2}(2\theta
_{12})$, the minimum of $\chi ^{2}$ moves to the left for larger values, or
to the right and on into positive values of $K_{-}$ for smaller values of $%
\sin ^{2}(2\theta _{12})$. This feature is a result of the interplay between
the two terms in Eq. (\ref{eq:P21new}), where increases in $\sin
^{2}(2\theta _{12})$ must be compensated by decreases in $K_{-}$ to keep the
minimum $\chi ^{2}$ at zero. Table \ref{tab:s2th12} gives the location in
the $K_{-}$ variable and values of the corresponding minimum of $\chi ^{2}$
for selected values of $\sin ^{2}(2\theta _{12})$ in the range $0.857\pm
0.024$. Parameter values are $K_{+}=0, a=0.03\ $and $b=0$, and $\sin ^{2}(2\theta _{13})_{eff}=0.089$%
, but fits have very little sensitivity to these choices. 
\begin{table}[tbph]
\begin{center}
\renewcommand{\arraystretch}{1.5} 
\begin{tabular}{|l|l|c|c|c|c|c|c|c|}
\hline
$\sin^2(2\theta_{12})$ & 0.881 & 0.873 & 0.865 & 0.857 & 0.849 & 0.841 & 
0.833 &  \\ \hline\hline
$K_-|_{min}$ & -0.0259 & -0.0168 & -0.0082 & 0.0 & 0.0078 & 0.0153 & 0.0225
&  \\ \hline
$\chi^2_{min}$ & 2 $10^{-4}$ & 3 $10^{-5}$ & 5 $10^{-5}$ & 0.0 & 2 $10^{-5}$
& 2 $10^{-5}$ & 8 $10^{-5}$ &  \\ \hline
"penalty" & 1.0 & 0.44 & 0.11 & 0.0 & 0.11 & 0.44 & 1.0 &  \\ \hline
\end{tabular}
\renewcommand{\arraystretch}{1}
\end{center}
\caption{At selected values of $\sin ^{2}(2\protect\theta _{12})$ within its 
$1\protect\sigma $ range, we list the values of the variable $K_{-}$ at the
minimum $\protect\chi ^{2}$, with its value, for the test of the NSI NH fit
to SMM "data" with central values for all its input parameters. The bottom
row lists the value of the additive "penalty" term $((\sin ^{2}(2\protect%
\theta _{12})-0.857)/0.024)^{2}$ understood in the $\protect\chi ^{2}$
definition, Eq. (\protect\ref{eq:chisq}). Energy detection statistical and
systematic error parameters are chosen at $a=0.03$ and $b=0$.}
\label{tab:s2th12}
\end{table}
The "penalty" term added to values $\chi ^{2}$ gives the weight against
values of $\sin ^{2}(2\theta _{12})$ away from the central value. This
weighting breaks the degeneracy between the variables $\sin ^{2}(2\theta
_{12})$ and $K_{-}$. It guarantees that values of $K_{-}$ "hiding behind" $%
\sin ^{2}(2\theta _{12})$ become less and less likely as deviations from the
central value grow. \emph{The size of values of $K_{-}$ are thereby limited
in the same measure as the digressions of $\sin ^{2}(2\theta _{12})$ away
from its central input value.}

This exercise demonstrates that a confusion effect arises in the analysis of
medium baseline experiments. A perfectly good fit to data by the SMM with
preferred input parameters can be mimicked by a NSI fitting model with a
choice of $\sin ^{2}(2\theta _{12})$ that is different from the SMM "data"
choice, making the procedure of constraining input parameters by using the
reactor neutrino data alone seriously model dependent. However, the smaller
the uncertainties in the input parameter $\sin^2(2\theta_{12})$ that are determined from other,
independent experiments with different physics at the detector and combined with
the medium baseline reactor analysis, the smaller the allowed
range of values of NSI parameters from zero.  The combined
data becomes more sensitive to the presence of NSI. 
In the example
above, values of $|K_{-}|~\geq 0.026$ are ruled out at $1\sigma $. With the
conventional one degree of freedom connection between confidence level (C.L.)
and {$\chi ^{2}$, $|K_{-}|\geq 0.04$ is ruled out at $90\%$ C.L., \emph{which
is essentially the same as the currently available bounds. Improvements in
uncertainty in $\sin ^{2}(2\theta _{12})$ in future experiments like SNO+ 
\cite{sno+} and Hyper-Kamiokande \cite{hyperK}
will improve the sensitivity of medium baseline reactor experiments to the $%
K_{-}$ parameter by the same factor as the improvement in input parameter
precision.}

As pointed out earlier, whether the distinction between the NH and IH cases
is stronger or weaker when NSI are present than it is in the SMM case
depends upon the sign of $K_{-}$. This gives another handle on the detailed
dependence on $K_{-}$ and $K_{+}$, and we investigate the sensitivity of the
MH determination from medium baseline data to these parameters.

\section{NH vs. IH as a function of the NSI parameters $K_{+}$ and $K_{-}$}

Our measure of the distinction between the spectrum expected in the NH mass
splitting case vs. the spectrum in the IH case is in the form of a $\chi ^{2}
$ that assumes data given by one case and the fit attempted with the other 
\cite{ghot}. Statistical and systematic uncertainties, as well as
uncertainties of the input parameters, will blur the contrast between the
two MH possibilities. Assuming that the IH is the data, for example, we
write the corresponding $\chi ^{2}$ as 
\begin{equation}
\Delta \chi _{MH}^{2}=\Sigma _{i=1}^{N}\left( \frac{\frac{dN}{dE_{\nu }}%
^{NH}-\frac{dN}{dE_{\nu }}^{IH}}{\surd \frac{dN}{dE_{\nu }}^{IH}}\right)
_{i}^{2}(\Delta E_{\nu })_{i},  \label{eq:chisqMH}
\end{equation}%
where $(\Delta E_{\nu })_{i}$ is the width of the $i^{th}$ energy bin. It is
understood that the NH and IH rates depend on the NSI parameters $K_{+}$ and 
$K_{-}$ (which can be taken as zero to regain the standard mixing result) as
well as the standard 3-neutrino mixing angles $\theta _{ij}$, mass-squared
differences $\Delta m_{21}^{2}$ and $\Delta m_{31}^{2}$ and the baseline $L$%
. Again, the summation is over the "data bins", which we take every $0.01$
MeV between $1.8$ and $8$ MeV. Using Eq. (\ref{eq:chisqMH}), we look at the
effects of varying the NSI parameters, effectively marginalizing only over
the range of input parameter values, $\Delta m_{31}^{2}$, which has the only
significant impact on the value of $\chi ^{2}$, as pointed out in \cite{ghot}%
.

In Fig. \ref{fig:chisqMHmin} we show the $\Delta \chi _{MH}^{2}$ landscape
produced by Eq. \ref{eq:chisqMH} for the value $\Delta
m_{31}^{2}=2.343\times 10^{-3}$, where $\Delta \chi _{MH}^{2}$ takes its
minimum value when NSI are not present . The figure is normalized by $\Delta
\chi _{MH}^{2}$ for $K_{-}=K_{+}=0,$ the SMM case. The baseline is $50$ km,
central values are chosen for all input parameters except $\Delta m_{31}^{2}$%
, and the energy uncertainty parameters are taken for illustration to be $%
a=0.03$ and $b=0$, which are among the values explored in \cite{ghot}. The
study presented here is the most straightforward generalization to NSI of
their analysis. 
\begin{figure}[tbph]
\begin{center}
\includegraphics[width=4in]{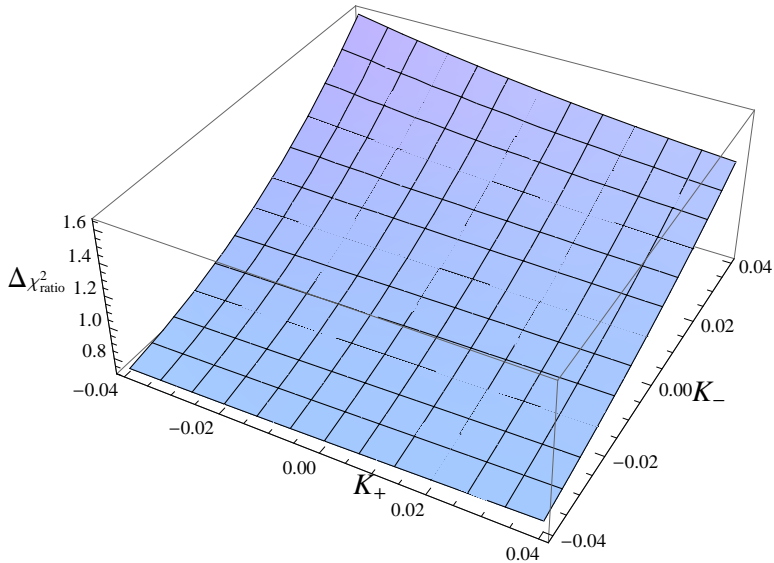}
\end{center}
\caption{For $K_{+}$ and $K_{-}$ ranging between -0.04 and +0.04, we show
the values of $\Delta \protect\chi _{MH}^{2}(K_{+},K_{-})/\Delta \protect%
\chi _{MH}^{2}(0,0)$ at the value of $\Delta m_{31}^{2}$ where $\Delta 
\protect\chi _{MH}^{2}$ is a minimum. The case where IH is assumed to be the
"data", as computed in Eq.(\protect\ref{eq:chisqMH}), is used for the figure.
}
\label{fig:chisqMHmin}
\end{figure}
As explained in Sec. III.B.2, for any value of $\Delta m_{31}^{2}$ the
smallest difference between NH and IH should occur where $P_{32}$ is
smallest and the largest difference should occur when $P_{32}$ is largest.
This is qualitatively reflected in the behavior shown in Fig. \ref%
{fig:chisqMHmin}. The features of the plots as they depend on $K_{+}$ and $%
K_{-}$ are explained in detail in Sec. III.B.2. For any value of $K_{+}$%
, the sensitivity to the MH as measured by $\Delta \chi ^{2}$ is greater in
the presence of NSI if $K_{-}>0$ and less if $K_{-}<0$. When $K_{-}<0$, the
dependence on $K_{+}$ is very weak, it is completely negligible when $K_{-}=0
$ and becomes somewhat stronger when $K_{-}>0$. For convenience, in Fig. \ref%
{fig:chisqMHmin} we use the common value $\Delta
m_{31}^{2}=(2.32+0.023=2.343)\times 10^{-3}eV^{2}$, the value at the minimum
when NSI are zero, for all the points. In fact the minimum value of $\Delta
\chi ^{2}$ shows some dependence on $K_{+}$ and $K_{-}$, though the
qualitative features of the landscape shown in the figure do not change.

One can make a correspondence between the $\Delta \chi ^{2}$ values at fixed 
$a=0.03$ and $b=0$ energy uncertainties while NSI parameters $K_{+}$ and $%
K_{-}$ vary, and the values that $\Delta \chi ^{2}$ takes on in the$\
K_{+}=K_{-}=0$ SMM case while $a$ and $b$ vary. For example, with $K_{+}$ in
the range from $0.0-0.02$,$\ K_{-}=0.04$ and $a=0.03$ and $b=0$, the $\Delta
\chi ^{2}\simeq 6.2$ is about the same as the IH $\Delta \chi _{min}^{2}$
value for $a=0.02$, $b=0.01$ at $50$ km in Fig. $7$ of \cite{ghot}, compared
to the $\Delta \chi ^{2}\simeq 3.5$ shown in Fig. $6$ of \cite{ghot} for $%
a=0.03$ and $b=0$. This improved sensitivity is driven by the positive value
of $K_{-}$, as indicated in our Fig. \ref{fig:chisqMHmin}. Conversely, as $%
K_{-}$ takes on negative values, the sensitivity is degraded. Combined with
hints of NSI in fits to the event spectrum or hints from other, independent
experiments, an anomalously low or high sensitivity to the MH may indicate
that NSI are at work.

We do not pursue this in detail here, since our limited goal is to identify
points where planned medium baseline reactor neutrino precision experiments
can be sensitive to NSI and points where NSI mimic the effects of
uncertainties in the value of input parameters. The latter degeneracies make
the precision determination of the MH and of SMM parameters model dependent.

\section{Summary and conclusions}

Current limits on NSI parameters are tight enough and the recently measured
value of reactor $\bar{\nu _{e}}$ disappearance probability is large enough
that NSI effects can be analyzed transparently at leading order in the
parameters. In this framework, we identified two effective NSI parameters
that play a role in short and medium baseline neutrino disappearance
analysis, $K_{+}$ and $K_{-}$, which are combinations of the complex flavor
violating NSI coefficients and, in the case of $K_{+}$, the standard mixing
matrix CP$-$violating phase $\delta $. The short baseline experiments \cite%
{dchooz, db, reno}, constrain a combination of $\sin ^{2}(2\theta _{13})$
and $K_{+}$, and we used this constraint as an essential input to our
analysis of consequences for medium baseline $\bar{\nu _{e}}$ disappearance
experiments that are in the planning stage, aimed at determining the
neutrino MH and the precision measurement of mixing angles and magnitudes of
mass-squared differences.

In Sec.III.A we outlined the constraint linking $K_{+}$ and $\sin
^{2}(2\theta _{13})$, essentially reducing the analysis of the medium
baseline experiments to a two parameter problem, either $K_{+}$ and $K_{-}$
or $\sin ^{2}(2\theta _{13})$ and $K_{-}$. The constraint is shown in Fig. 
\ref{fig:constrain1}. In Sec.III.B, we showed that the event rate
spectra at fixed baseline,$\ 50$ km being the default value throughout our
analysis, are sensitive to the NSI parameters for given, fixed values of $%
\sin ^{2}(2\theta _{12})$ and $\sin ^{2}(2\theta _{13})$, even with generous
energy uncertainties allowed. When $\sin ^{2}(2\theta _{12})$ was varied
within its $1\sigma $ range with no NSI present, we found the effects were
mimicked by NSI parameters varied in the same range while $\sin ^{2}(2\theta
_{12})$ was held fixed at its central value, Figs. \ref{fig:rates0} and \ref
{fig:rates6}. Outside this range, the variation in NSI starts to show up, as
we show in Fig. \ref{fig:envelopes}. This degeneracy was then explored
further in Sec.IV. In Sec.III.B.2, the long oscillation length term,
sensitive to the probability factor $P_{21}$, was determined to depend only
on $\sin ^{2}(2\theta _{12})$ and $K_{-}$, the dependence on $K_{+}$ being
entirely negligible, simplifying the statistical analysis in
Sec.IV. The short oscillation length, MH sensitive term is controlled by
the probability factor $P_{32}$, which we showed depends linearly on $K_{-}$%
, with a coefficient that depends only weakly on $K_{+}$ through its
constraint with $\sin ^{2}(2\theta _{13})$.  The impact of NSI is not
as dramatic on the MH sensitive terms as on the scale of the rate spectrum,
as indicated by comparing Figs. \ref{fig:rates6}, left panel, and \ref%
{fig:difNHIH}, right panel. Nonetheless, with sufficient statistics and reduction
of systematic uncertainties, the distinction can be made \cite{ghot}. However,  the effect 
of the NSI is degenerate with variation in $\sin ^{2}(2\theta _{13})_{eff}$, 
emphasizing the need for precision determination of $\sin^2(2\theta_{13})$ 
independent of reactor neutrino experiments \cite{solaccel}.

Following up on these considerations in Sec.IV.A, we varied the value of $%
\sin ^{2}(2\theta _{12})$ within its $1\sigma $ range in the NSI fit
function, while adopting as "data" the SMM with central values of input
parameters. Modulo the "pull" term, which penalizes the fit by one unit of $\chi^2$
at 1$\sigma$, 4 units at 2$\sigma$ and so forth, we found a continuum 
of very nearly degenerate $\chi ^{2}$ minima with values
typically of the order $10^{-4}$ to $10^{-5}$. The spread
in $K_{-}$ values at the minima is roughly from $-0.025$ to $+0.025$ as $%
\sin ^{2}(2\theta _{12})$ runs over its 1 $\sigma$ uncertainty, $0.857\pm 0.024$. The
effects of varying other parameters were negligible.  The bounds on 
the $K_{-}$ parameter follow the bounds on the input uncertainties
in $\sin ^{2}(2\theta _{12})$ quite closely. This observation has important
consequences: the tighter the input uncertainties, the higher the
sensitivity to the NSI parameter $K_{-}$, so doubling the precision doubles
the sensitivity to NSI parameters!

Finally, in Sec.V we investigated the influence of NSI on the statistical
discrimination between NH and IH in medium baseline experiments.
Generalizing the pattern of analysis of Ref. \cite{ghot}, we showed that,
for a given statistical and systematic energy determination uncertainties,
the allowed NSI parameter ranges can either enhance or suppress the
sensitivity expectations based on the SMM analysis. Improved, independent
determination of $\sin ^{2}(2\theta _{13})$ can \emph{improve the sensitivity to
NSI parameter $K_-$ }in the MH determination in medium baseline experiments, 
as remarked in Sec.III.B.1. Moreover, it can \emph{provide sensitivity to the parameter $K_+$} 
in the short baseline experiments.

We conclude that the planned medium baseline, reactor neutrino experiments
to determine the MH and to make precision measurements of neutrino parameters 
are also good probes of NSI as the measurements of $%
\sin ^{2}(2\theta _{12})$ and $\sin ^{2}(2\theta _{13})$ in independent
experiments such as SNO+ \cite{sno+}, Hyper-Kamiokande \cite{hyperK} 
and T2K \cite{T2K,T2K2} become more precise.
This will enable one to restrict the degeneracies pointed out in Secs. III
and IV and increase sensitivity to NSI \cite{solaccel}. In particular, the overall
spectrum statistics provide a method for detecting the presence of NSI at
levels below those available in the literature. This diagnostic is not
sensitive to the sign of the relevant parameter. If there is evidence from
the gross spectral features that NSI are present, the short oscillation
length pattern that reveals the MH also provides additional information on
its sign, which determines whether the MH signal is suppressed or enhanced.

\begin{acknowledgments}
D. W. M would like to thank the Physics Department of CIIT Islamabad for
hosting his visit while this work was begun. The visit was supported by the
Higher Education Commission (HEC) of Pakistan. A. N. K. would like to thank
The University of Kansas (KU) Particle Theory Group for kind hospitality
during his visit to KU, the HEC for support for his graduate studies and for
his visit to KU under the Indigenous Ph.D. Fellowship Program (5000
Fellowships) Batch-IV and International Research Support Initiative Program.
The authors benefitted from detailed communications with Dr. Shao-Feng Ge
and Dr. Yoshitaro Takaesu. D.W. M. and A. N. K received support from DOE
Grant. No. De-FG02-04ER41308.
\end{acknowledgments}


\begin{thebibliography}{99}
\bibitem{solaccel} Analysis of solar experiments including the NSI at the source ($^8$B decay for instance) but different NSI at the detector, or no NSI at the detector, combined with the current analysis would be necessary.  The neutrino detection in the purely leptonic or neutral current modes would satisfy this requirement. To the best of our knowledge, only NSI in propagation and/or detection have been done for the solar neutrino case. See, for example \cite{to} and references therein. This is a topic for further research and goes beyond the scope of this paper.
\bibitem{fn1} A more general form with flavor dependent right-handed coefficients is given in \cite{bbf-m}. We simplify to left-handed helicity in the work here, so our $K_{ij}$ are equivalent to $\epsilon^{udL}_{ij}$ in \cite{bbf-m}.
\bibitem{fn2} In any case, the normalization is absorbed into a redefinition of the
effective Fermi constant, whose experimental constraints via universality
show up in the model independent bounds on the NSI parameters \cite{bbf-m}
we use here.
\bibitem{fn3} Note that the $P_{ij}$ notation used in \cite{zwcw} absorbs the factors $%
\sin ^{2}x_{ij}$ as well.
\bibitem{fn4} The sign and value of $x_{21}$ have been determined by solar neutrino experiments.
\bibitem{fn5} As remarked earlier, we assume that phase "accidents" $\cos (\delta +\phi
_{e\mu })=\cos (\delta +\phi _{e\tau })=0$ or $\cos \phi _{e\mu }=\cos \phi
_{e\tau }=0$ don't happen.
\bibitem{fn6} This result is essentially the same as that of Eq.$~(28)$ in \cite{klos} for
the case of same physics at source and detector, and as a linear
approximation to Eqs.$(17)$ and $(18)$ of \cite{oz} or Eqs. $(3.1)$ and $%
(3.2)$ of \cite{lmrz}. A comparison between our Fig. \ref{fig:constrain1}
and the right-hand-side of Fig. $1$ of \cite{oz} can be made roughly by
fixing $\tilde{\theta}_{13}$ at $8.5$ degrees, which corresponds to $\sin
^{2}(\theta _{13})_{eff}=0.089$, and comparing the range of $\theta _{13}$
values for $|\epsilon |~=0.01$ to that corresponding to difference between
its value at $K_{+}=-0.01$ and $+0.01$ in our Fig. \ref{fig:constrain1},
which is between about $8$ degrees and $9$ degrees. There is agreement
between the two figures, as far as we can tell from Fig. $1$ in \cite{oz}.
\bibitem{fn7} Unless we specify otherwise, we will assume maximal 2-3 mixing.
\bibitem{fn8} Varying the value of $\sin^2(2\theta_{13})_{eff}$ within its 1 $\sigma$ range 
has barely noticeable effect on the spectral curves - of the order of the widths of the
 lines in the figure.
\bibitem{fn9}  In the extreme case of very large NSI
parameter values, the overall sign of $P_{32}$ can change and reverse the
expected pattern of NH vs. IH oscillations. Within the constrained range
of parameters, roughly where $|K_{\pm }|~\leq 0.05$, there is no point where
it becomes zero or negative, so the NSI cannot reverse the expected signs of
NH and IH. At the point $(-0.1,-0.2),$ more than $4$ standard deviations
outside the constrained range for the NSI parameters and outside of the
linear region, $P_{32}=0$, and it becomes negative for values of $K_{-}<-0.2$
with $K_{+}\leq -0.1$.
\end{thebibliography}
\end{document}